\documentclass{article}
\usepackage{graphicx}
\usepackage{epsfig}
\usepackage{amsfonts}
 \usepackage{amssymb}
\usepackage[dvips]{color}
\date{\today}

\newcommand{\insertplot}[5]{\begin{figure}
 \hfill\hbox to 0.05in{\vbox to #5in{\vfill
 \inputplot{#1}{#4}{#5}}\hfill}
 \hfill\vspace{-.1in}
 \caption{#2}\label{#3}
 \end{figure}}
 \newcommand{\inputplot}[3]{
 \special{ps: plotfile #1}
}
\newcounter{fig}

\renewcommand{\a}{\alpha}

\newcommand{\ee}{\end{equation}}
\newcommand{\eea}{\end{eqnarray}}
\newcommand{\be}{\begin{equation}}
\newcommand{\bea}{\begin{eqnarray}}

\begin{document}

 \title{ 
Scalarized-charged wormholes in Einstein-Gauss-Bonnet gravity
} 

\author{
{\large Y. Brihaye}$^{\dagger}$, 
{\large J. Renaux} $^{\dagger}$
\\ 
\\
$^{\dagger}${\small D\'ep. {\it Physique de l'Univers, Champs et Gravitation}, Universit\'e de
Mons, Mons, Belgium}
}
\maketitle 
\begin{abstract} 
The Einstein-Maxwell-Klein-Gordon Lagrangian is supplemented by a non-minimal coupling of the real scalar field
to the  Gauss-Bonnet  invariant. The non minimal coupling function
is chosen as a general second degree polynomial in the scalar field for which the system is known to admit
hairy black holes. The new interaction 
leads naturally to a violation of the null energy condition, allowing for wormholes to exist without the need
of exotic matter.  
 Spherically symmetric, charged wormholes are constructed and their domain of existence is determined in terms 
of the different choices of the non-minimal coupling constants and of the electric charge. 
A special emphasis is set to the case of the purely quadratic coupling function.
 A phenomenon reminiscent to spontaneously scalarised black holes occurs for wormholes. 
The interaction with the electromagnetic field  leads to new families of wormholes supported by a non-vanishing,
large enough,
electric charge.   
\end{abstract} 
\section{Introduction}
In the last decades the Gauss-Bonnet (GB) term has played an important role in the 
construction of several extensions of General Relativity (GR).
Apart from the mathematical aspects, the interest for this geometric invariant (which, by itself is a total divergence in four space-time dimensions) is related to the fact that it emerges
in effective theories describing the low energy limit of some string theories \cite{ref1},\cite{ref2}. 
It is a coupling involving some extra field -- e.g. a scalar field like the dilaton -- that leads to a non-trivial
gravitational interaction.
The general  theory extending gravity by mean of a scalar field was elaborated by Horndeski \cite{horndeski}.
This theory is very rich and characterized by a Lagrangian density containing  large arbitrariness
 (see e.g. \cite{Maselli:2015yva}).
A few specific truncations of the theory were considered by several authors. One of them
consists in selecting a non minimal term of the form $H(\phi) {\cal L}_{GB}$ where 
the scalar field $\phi$ interacts with geometry through   
the Gauss-Bonnet term ${\cal L}_{GB}$ and a function $H(\phi)$.
\\
Apart from cosmological implications,
one interesting issue of tensor-scalar gravities is that they allow for new kinds of compact objects to exist, 
e.g. hairy black holes and wormholes, see e.g. \cite{Herdeiro:2015waa},\cite{Sotiriou:2015pka},\cite{Volkov:2016ehx} for recent reviews. 
One example is the family of hairy black holes  obtained in \cite{Herdeiro:2014goa} with General Relativity  minimally coupled to a massive, complex scalar field.
In this case, the No-hair theorems for black holes related to scalar hair \cite{no_hole_old,no_hole_new} are evaded by constructing a 
rotating black hole and  the synchronisation of the spin of the black hole with the angular frequency of the scalar field.
\\
\par Recently Horndeski theory was been studied thoroughly in the context of  Galileon theory \cite{galileon}
and some generalizations  of the latter  \cite{Deffayet:2011gz}.
Galileon theory with a shift symmetric scalar field was considered in
\cite{Sotiriou:2013qea} and still leads to a large family of models. 
Considering $H(\phi)$ to be a linear function, 
hairy black holes have been constructed numerically and perturbatively \cite{Sotiriou:2014pfa}. 
Dropping the requirement of shift symmetry, a coupling function of the form $H(\phi)= \gamma \phi^2$ 
 has been used in \cite{Silva:2017uqg}, \cite{Antoniou:2017acq}
(and  some other forms in \cite{Doneva:2017bvd},  \cite{Antoniou:2017hxj}). 
In these latter models, the existence of hairy black holes
results from an unstable mode associated to the linearized equation of the scalar field
in the background of a Schwarschild geometry  and sourced by the non-minimal coupling term. 
The coupling constant $\gamma$ plays the role of a spectral parameter of the linear equation 
and black holes get
{\it spontaneous scalarized}   at  a critical value of this coupling constant.
\\
\par
Next to black holes, another interesting class of solutions appearing in gravity theories are wormholes first discussed in \cite{wormhole_1} and interpreted in \cite{wormhole_2}.
It is well known that viable (meaning: traversable) wormholes need the existence of exotic matter, i.e. matter that violates the 
null energy condition \cite{wormhole_3}. However, as has been demonstrated in \cite{Kanti:1995vq}, the energy-momentum tensor
associated to a  GB term violates this energy condition, opening the possibility for constructing wormholes
in Einstein-Gauss-Bonnet models coupled to {\it normal} matter fields.
Such objects where indeed constructed in 
\cite{Kanti:2011jz,Kanti:2011yv} and in \cite{Antoniou:2019awm} where the properties, the
domain of existence and the stability have been studied in details for some choices of the coupling function $H(\phi)$.
\\
\par
In the present paper we will consider a scalar-tensor gravity model with $H(\phi) = \alpha \phi + \gamma \phi^2$
where $\alpha$, $\gamma$ are independent constants. With such a combination, used first in \cite{Brihaye:2018grv}, the features of the model containing 
spontaneously scalarised black holes  (for $\alpha=0$, $\gamma\neq 0$) and of the model
containing shift symmetry (for $\alpha\neq 0$, $\gamma=0$) appear together. 
Following \cite{Brihaye:2019kvj}, we supplement  the model by a Maxwell term  in order to study 
the influence of an electromagnetic field on the solutions.
In the following, we will present strong numerical evidences  that  wormholes exist in the model
and study how the presence of an electric potential affects their pattern.  
Our results reveal in particular that the presence of an electric field leads to charged wormholes  for both signs of the
coupling constant $\gamma$.  Only a subset of these solutions persists while the electric is suppressed progressively.
The paper is organized as follow: in section 2 we present the model, discuss the Ansatz and the boundary conditions.
Section 3 contains the presentation of several properties of the wormholes and
the determination of the 
domain of existence of these solutions in terms of the charge and of the coupling constants of the model. 
A summary of the results and perspectives are given in section 4. 
\\
\section{The model}
We are interested in wormhole-solutions associated with  Einstein-Maxwell-Klein-Gordon lagrangian extended by a non-minimal
coupling. The action considered is  the form
\be
   S = \int d^4 x \sqrt{-g} \bigg[ \frac{1}{16 \pi G} R - \nabla_{\mu} \phi \nabla^{\mu} \phi 
	- \frac{1}{4} F_{\mu \nu}F^{\mu \nu} + H(\phi) {\cal I}(g)  \bigg]
\label{lagrangian}
\ee
where $R$ is the Ricci scalar (in the following we will pose $8 \pi G = 1$),
 $F_{\mu \nu}$ in the electromagnetic field strength 
 and $\phi$ represents a real scalar field. 
The gravity sector is  supplemented by a non-minimal coupling of the scalar field to a geometrical invariant ${\cal I}(g)$.
In this paper, we choose this invariant  as the Gauss-Bonnet-scalar~:
  $${\cal I}={\mathcal L}_{GB}\equiv R^2 - 4 R_{ab}R^{ab} 
            + R_{abcd}R^{abcd} \ \ .$$ 
This combination  is well known to be a total derivatives in four dimensions but it contributes non trivially 
to the equations of motion through its interaction with the scalar field.

Several forms of the function $H(\phi)$ have been emphasized in the litterature to construct hairy black holes
and/or neutron stars in scalar tensor gravity.
 The purely linear case $H(\phi)=\alpha \phi$ corresponds to a shift-symmetric theory
adressed   in
 \cite{Sotiriou:2013qea}. A quadratic choice $H(\phi)=\gamma \phi^2$   was  considered in 
\cite{Silva:2017uqg} and 
\cite{Antoniou:2017acq}. Several other choices  of the function $H(\phi)$,
(e.g. $H(\phi) = \gamma \exp(-\phi^2)$)
have been used in \cite{Doneva:2017bvd},  \cite{Antoniou:2017hxj}.
Wormholes were constructed in \cite{Kanti:2011yv}\cite{Kanti:2011jz}
for a dilaton coupling function $H(\phi)= \alpha \exp(-\phi)$ and in \cite{Antoniou:2019awm}
for several monomials forms of the function $H(\phi)$.

In this paper, we choose
\be
\label{coupling}
               H(\phi) = \alpha \phi + \gamma \phi^2
\ee
where $\alpha$, $\gamma$ are independent coupling constants (without loosing generality we can assume $\alpha > 0$).
The choice (\ref{coupling}) can be considered as a truncation of a general analytic coupling function; in addition  it leaves  
the possibility to construct wormholes in the two limits  $\alpha=0$ and $\gamma=0$ 
where hairy black holes are known to exist (as explained in the introduction).
The behavior of the solutions with  the mixed coupling will be studied
as well as the influence of the electromagnetic field.
Let us add that the construction of hairy black holes with this combination (\ref{coupling}) 
was adressed  in  \cite{Brihaye:2018grv}.

\subsection{Ansatz}
We will be interested in spherically symmetric solutions and adopt  a  metric of the form
\be
     ds^2 = - f(r) a^2(r) dt^2 + \frac{1}{f(r)} dr^2 + (r^2+ r_0^2) d \Omega_2^2 
\ee
(note the change of notation: $g_{rr} = 1/f(r)$ with respect to \cite{Kanti:2011yv}).
This is completed by the scalar field and vector fields  of the form 
\be
\phi(x^{\mu}) = \phi(r) \ \ , \ \  
 A_0(x^{\mu}) = V(r) \ \ , \ \ A_1 = A_2 = A_3 = 0 .
\ee
Substituting the ansatz in the field equations, the  system can be reduced to a set of three  non linear differential
equations (plus a constaint) in the functions $f(r),a(r)$ and $\phi(r)$.
The potential $V(r)$ is easily eliminated by using the Maxwell equation, leading to 
\be
V'(r) = Q a/(r^2+r_0^2)
\ee
where $Q$ is an  integration constant which, for instance, is proportional to the electric charge of the solution.
The final equations are of the first order for the functions $f(r)$, $a(r)$ and of the second order for $\phi(r)$;
they are to be solved on the interval  $r \in [0,\infty]$,  the throat of the wormhole corresponds to the limit $r\to 0$.
\subsection{Boundary conditions}
For a fixed couple of parameters $(\alpha, \gamma)$ and of the constant $Q$, 
four  conditions on the boundary need to be fixed to specify a solution.
Because we  look for localized, asymptotically flat solutions, we require
\be
       \a(r \to \infty) = 1 \ \ \ \ , \ \ \ \ \phi_(r \to \infty) = 0 \ \ .
\ee 
The equations are singular in the limit $r\to 0$, and obtaining regularity require a very specific relation
between the functions and their derivatives at $r=0$. To specify this relation (and for later use)
 it is convenient to write the Taylor expansion of the fields
\be
\label{taylor}
    f(r) = f_0 + f_1 r + o(r^2) \ \ , \ \ a(r) = a_0 + a_1 r + o(r^2) \ \ , \ \ \phi(r) = \phi_0 + \phi_1 r + o(r^2)
\ee
 The relevant regularity condition is~:
\be
\label{condition}
       \phi_1^2 = \frac{2(1-f_0) - Q^2 \bigl(1 + 16(1-f_0)(\alpha + 2 \gamma \phi_0) \bigr) }
			{16 f_0 \bigl( \gamma - 2 (1-f_0)(\alpha+ 2 \gamma \phi_0)^2 \bigr) }
\ee
%
In the limit $\alpha=0$ , $|\gamma| \ll 1$, the condition simplifies to $\phi_1^2 = (1-Q^2/2-f_0)/(8 f_0 \gamma)$.
We see that for $\gamma > 0$, the condition implies $f_0 < 1-Q^2/2$; by contrast, for $\gamma < 0$,
the parameter $f_0$ is not bounded by the regularity condition.

Finally, the fourth boundary condition (necessary to specify a boundary value problem)
 is  imposed by fixing by hand the value $f(0)$ (or alternatively $\phi(0)$). 
This value somehow serves as a control parameter;
as we will see in the next section its variation is generally  limited to a
specific interval with bounds depending on $\alpha,\gamma$ and of $Q$.  
 
The different coefficients in (\ref{taylor}) can be computed recursively  in terms of $f_0,a_0,\phi_0$,
the charge $Q$ and the constants $\alpha, \beta$; the final expressions are quite involved and not illuminating.

\subsection{Physical parameters}
Along  black holes, the wormholes  solutions   can
 be characterized by their mass $M$, the electromagnetic charge $Q$ (if $V \neq 0$) 
and a charge, say $D$, characterizing the scalar field.
These charges are related respectively to the asymptotic decay
of the functions $f(r)$, $V(r)$ and $\phi(r)$
\be
     f(r) = 1 - \frac{2M}{r}  + O(1/r^2) \ \ , \ \ 
		  V(r) = V_0 - \frac{Q}{r} + O(1/r^2) \ \ , \ \ 
		\phi(r) = - \frac{D}{r} + O(1/r^2) \ \ , a(r) = 1 - \frac{1+D^2}{2 r^2} + \ \ + O(1/r^3).
\ee
The area of the throat is given by $A_{th} = \pi r_0^2$ and the curvature at the throat, 
say $R_0$ is given by $R_0= r_0/f(0)$ (see e.g. \cite{Kanti:2011yv}) for more details). 
The surface gravity at the throat, say $\kappa$, and temperature $T_H$ are defined as in the case of black holes.
With our choice of the metric, we find $\kappa = (f'(0) a(0) + 2 f(0) a'(0))/2$ and
$T_H = \kappa /(2 \pi)$. In terms of the Taylor expansion (and setting $\alpha=0$ for simplicity), 
the surface gravity takes the form
\be
    \kappa = \frac{a_0(2 f_0 \phi_1^2 + 2 - Q^2)}{16 \phi_1(\alpha + 2 \gamma \phi_0)}
\ee

The classification of the solutions can be simplified by exploiting the following scale invariance
of the equations 
\be
      r \rightarrow \lambda r \ \ ,  \ \ r_0 \rightarrow \lambda r_0 \ \ , \ 
			\gamma  \rightarrow \lambda^2  \gamma  \ \ , \ \ \alpha \rightarrow \lambda^2 \alpha.
\ee
The throat radius  $r_0$ of the wormhole can therefore be normalized to one without loosing generality.
This normalization has been used throughout  the numerical construction. 
\section{Numerical results}
We now discuss the pattern of  solutions in the space of the parameters $\alpha, \gamma$ and $Q$.
On a suitable domain of this triplet, 
a branch of solutions exist which can be labeled by the value $f(0)$ 
(or alternatively by $\phi(0)$), 
Along \cite{Kanti:2011yv}, we find that the various branches exist on a finite interval of these parameters
and approach critical configurations of three types
\begin{itemize} 
\item {Type A :} The wormhole approaches a black hole as $f(0) \to 1$.
\item {Type B :} The limiting configuration presents a singularity at the throat.
\item {Type C :} The limiting configuration presents a singularity at an intermediate value $r_s$, with $0 < r_s < \infty$.
\end{itemize} 
\par Since the non-linear equations do not admit closed form solutions, we solved the system by 
using the numerical routine COLSYS \cite{COLSYS}.
It is based on a collocation method for boundary-value differential equations and on damped Newton-Raphson iterations.
The equations are solved with a mesh of a few hundred points and  relative errors of the order of $10^{-6}$.
The variation of  four independent quantities, namely $\alpha, \gamma, Q$  and $f(0)$,
renders the determination of the domain quite cumbersome;
we therefore limit our investigation to two sub-cases, namely~: 
\begin{itemize}
\item (i) We vary $\gamma, Q$ for the quadratic coupling (i.e. $\alpha=0$). The influence of the charge of 
          spontaneous scalarisation can then be appreciated.  
\item (ii)  Varying the constants $\alpha, \gamma$ in the uncharged, i.e. with $Q=0$. 
\end{itemize}


\subsection{Pure quadratic coupling} 

\subsubsection{Uncharged solutions}  We first discuss the case  $Q=0, \alpha = 0$. 
The corresponding theory admits a family of hairy black holes which bifurcates from the Schwarschild solution
at a critical value of the coupling constant $\gamma$. To see this, the
the Klein-Gordon equation sourced by the scalar-Gauss-Bonnet interaction term 
is considered in the Schwarschild background. It turns out that a regular, localized solution of the linear equation
exists for a specific value of the coupling constant $\gamma$, say for $\gamma_c \sim 0.1814$  
(solutions with nodes exist as well, with different values of $\gamma$, but are not studied here).
From the critical value, a branch solutions  of the system of coupled equations 
exist for $\gamma \in [0.172,0.1814]$; they have $\phi(r) \neq 0$ and then constitutes hairy
black holes \cite{Silva:2017uqg}. 
Similar solutions with other choices of the  function $H(\phi)$ were constructed in \cite{Antoniou:2017acq}).
\begin{figure}[h!]
\begin{center}
{\label{non_rot_cc_1}\includegraphics[width=10cm, angle = 0]{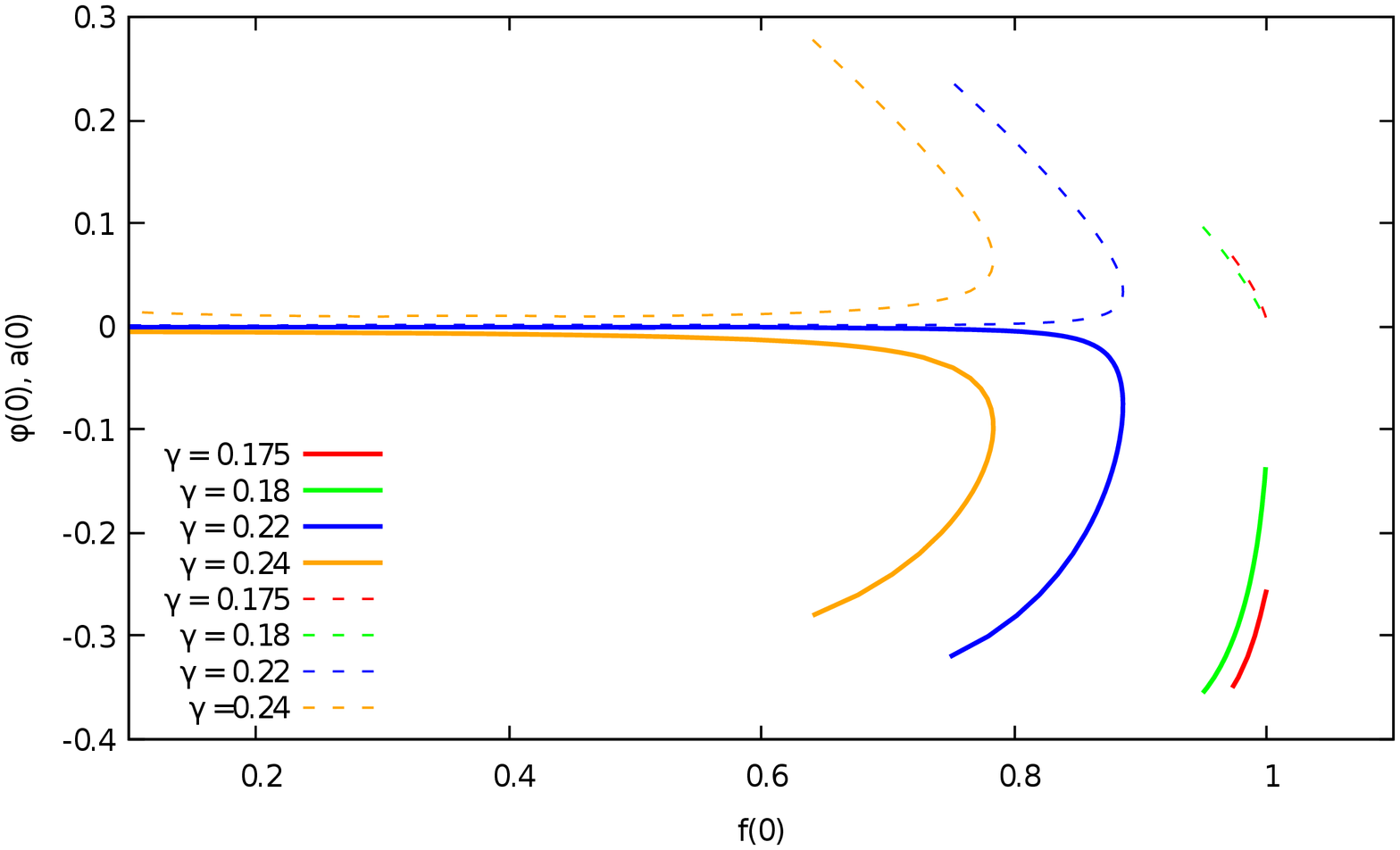}}
{\label{non_rot_cc_2}\includegraphics[width=10cm, angle = 0]{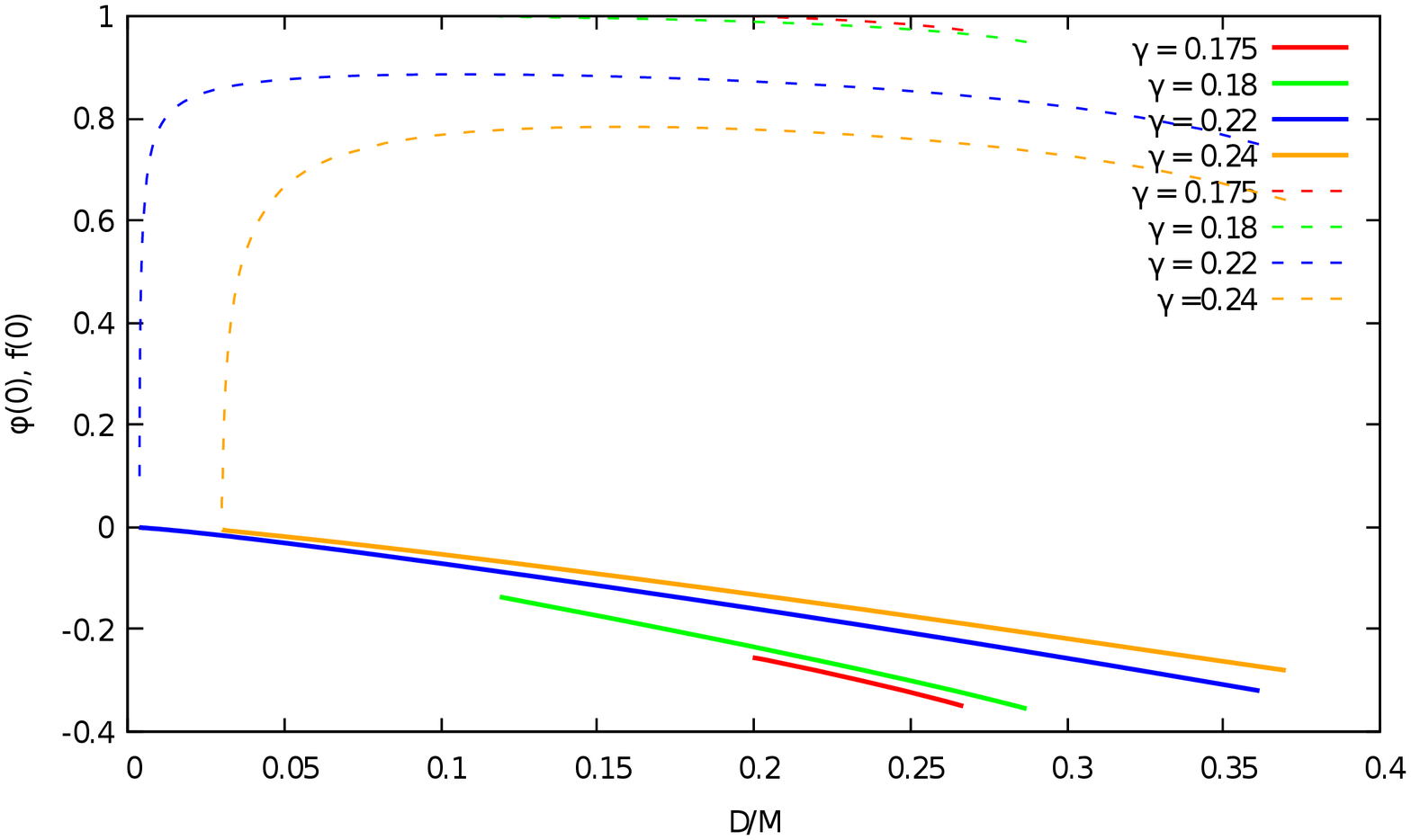}}
\end{center}
\caption{Up: The values of $a(0)$ and $\phi(0)$ (dashed and solid curves respectively)
as functions of $f(0)$ for several values of $\gamma$. 
Down: The values of $f(0)$ and of $\phi(0)$ ( dashed and solid curves respectively) as functions of 
the dimensionless parameter $D/M$.
\label{fig_ga_0}
}
\end{figure} 

Turning to  the wormhole equations reveals that solutions with $\phi(r) \neq 0$ 
exist  for $\gamma \gtrsim 0.172$ as well.
For a fixed value of $\gamma$ slightly larger than this threshold, a family of wormholes exists 
with a finite extend of the parameters $\phi(0)$ and  $f(0)$. 
For the discussion we use $f(0)$ as a control parameter, the counting of the branches of solutions 
will refer to $f(0)$. 

Two different patterns occur according to the value of $\gamma$ as revealed by Fig. \ref{fig_ga_0}.
 In each case  the phenomenon limiting the branches  has  different characteristics. 
\begin{itemize}
\item{(i)} For $\gamma \in [0.172, 0.1814]$ a single branch of solutions exist 
with $f(0) \in [c_1, 1]$ (the value $c_1$ depends on $\gamma$).
The solutions approach a hairy black holes for $f(0) \to 1$.   In the limit $f(0) \to c_1$, 
 a configuration presenting a singularity at some intermediate radius is approached. 
\item{(ii)} For $\gamma > 0.1814$ no hairy black hole that can be approached do exist. Instead
two branches of solutions exist respectively with $f(0) \in [0, c_2]$ (say branch I) and with $f(0) \in [c_1,c_2]$ (branch II).
The branch I ends up into a singular configuration at the throat since $f(0) \to 0$. The limit of branch II for $f(0) \to c_1$
is a configuration presenting a singularity at an intermediate radius. 
\end{itemize}
On the upper part of Fig.\ref{fig_ga_0} the dependance of $\phi(0)$ on $f(0)$ is shown by the solid lines
for several values of $\gamma$.
The transition between the one-branch  and the two-branches regimes clearly appears.
The corresponding values of $a(0)$ are plotted by the dashed lines. 
The same quantities are reported in terms of the dimensionless ratio $D/M$ on the lower part of Fig.\ref{fig_ga_0}.
The scalar function $\phi(r)$ is negative, presenting a local maximum at some intermediate radius and then increases,
explaining the positivity of the charge $D$.  
Completing, the data, Fig. \ref{kappa} (left side)  reveals that, when two branches are present,
the surface gravity $\kappa$ of the solutions of the main branch  depends only a little from $f(0)$;
by contrast $\kappa$ becomes large
and strongly dependent of $f(0)$ for the solutions of the second branch.

Let us point out that, like in Ref. \cite{Kanti:2011yv}, we managed to construct uncharged wormholes for $\gamma >0$ 
and found no evidence of solutions for $\gamma < 0$.

\subsubsection{Influence of the charge}
From the regularity condition (\ref{condition}), different patterns of charged solutions can be expected 
depending on the sign of the constant $\gamma$. It  turns out, indeed,
 that solving the equations for $Q \neq 0$ leads to new families of solutions that have no limit
for $Q \to 0$. 
\\
{\bf Positive $\gamma$.}
As pointed out already, in the absence of an electric potential (i.e. $Q=0$), 
wormholes can be constructed for positive values of $\gamma$ only. 
Increasing the charge parameter $Q$  progressively, these uncharged solutions
get continuously deformed leading to families of charged wormhole exist.
These exist up to a maximal value of the parameter, say $Q_c$, of the charge parameter.
In all cases that considered, 
the limiting configuration  presents a singular  geometry at the throat as the 
metric function $a(0)$ indeed approaches zero for $Q \to Q_c$.
Since these results are somehow expected, we do not present details and 
concentrate on the solutions available   for negative values of $\gamma$.
\\
\begin{figure}[h!]
\begin{center}
{\label{non_rot_cc_1}\includegraphics[width=8cm, angle = -0]{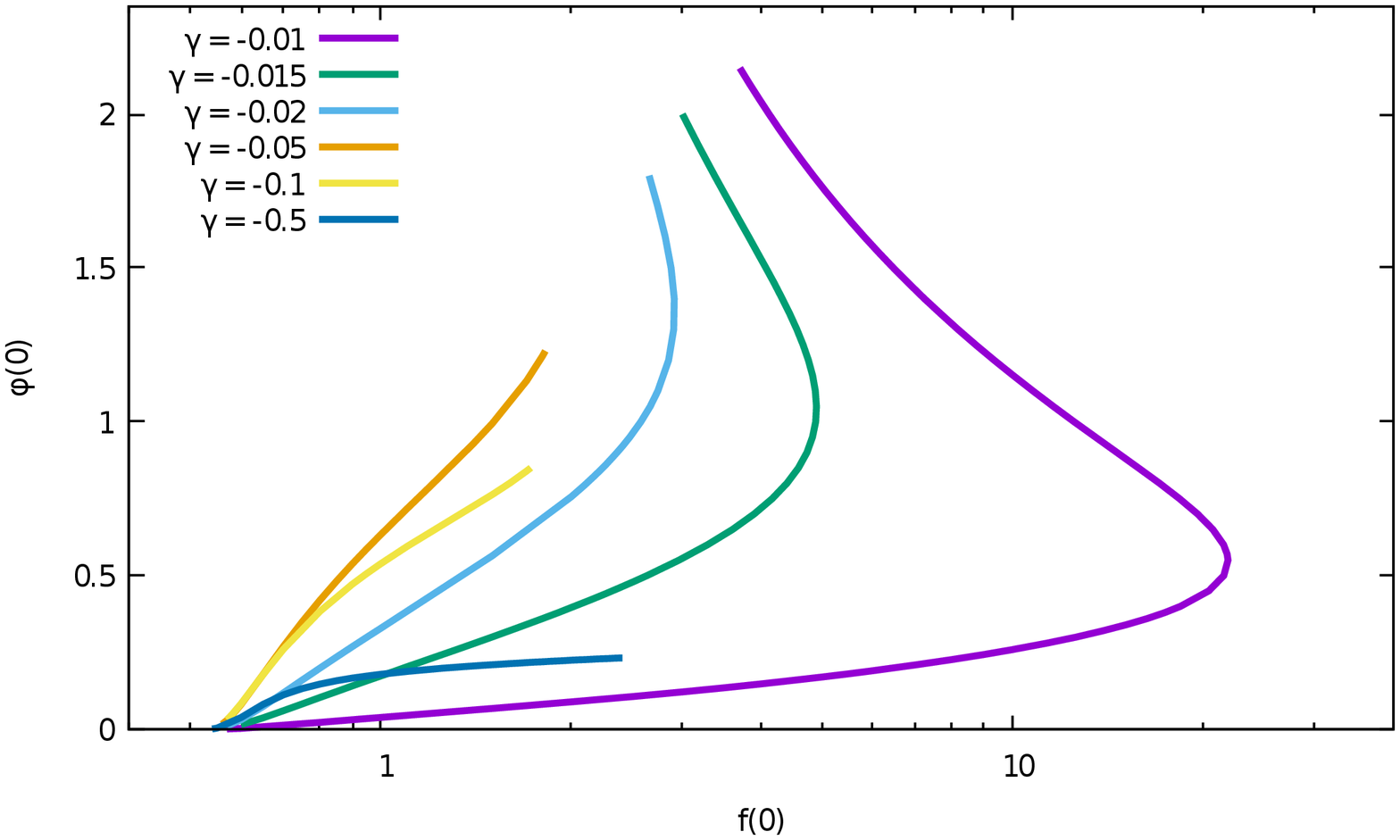}}
{\label{non_rot_cc_2}\includegraphics[width=8cm, angle = -0]{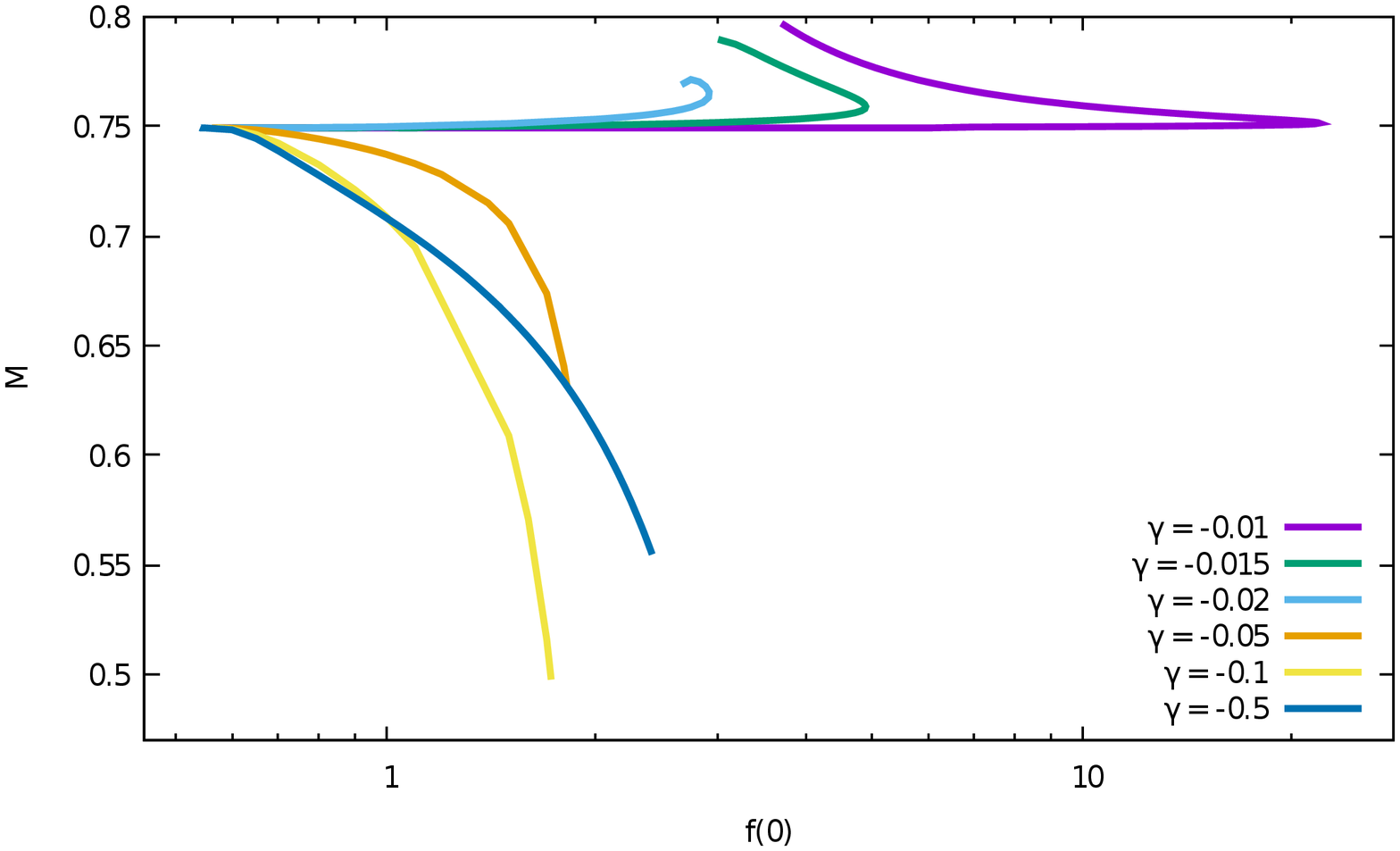}}
\end{center}
\caption{left: The value of $\phi(0)$ as function of $f(0)$ for several values of $\gamma$ and $Q=1$.
Right: Idem for the Mass. 
\label{data_ga2_neg_0}
}
\end{figure}
\\ 
{\bf Negative $\gamma$.}
Interestingly, the analysis of the equations with $Q \neq 0$
reveals that, for sufficiently large $Q$,  new branches of solutions exist for $\gamma < 0$.
\begin{figure}[h!]
\begin{center}
{\label{non_rot_cc_1}\includegraphics[width=8cm, angle = -0]{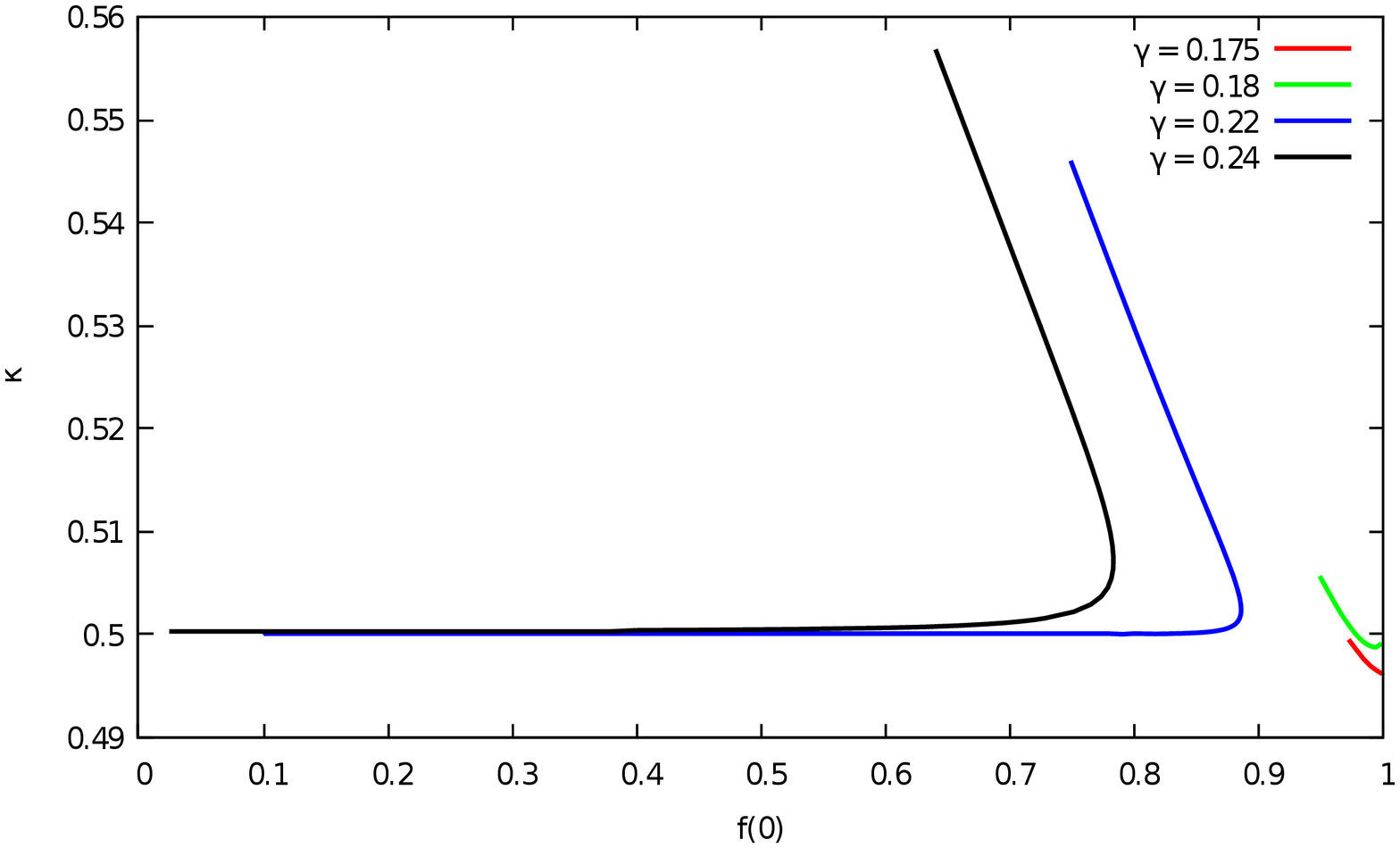}}
{\label{non_rot_cc_2}\includegraphics[width=8cm, angle = -0]{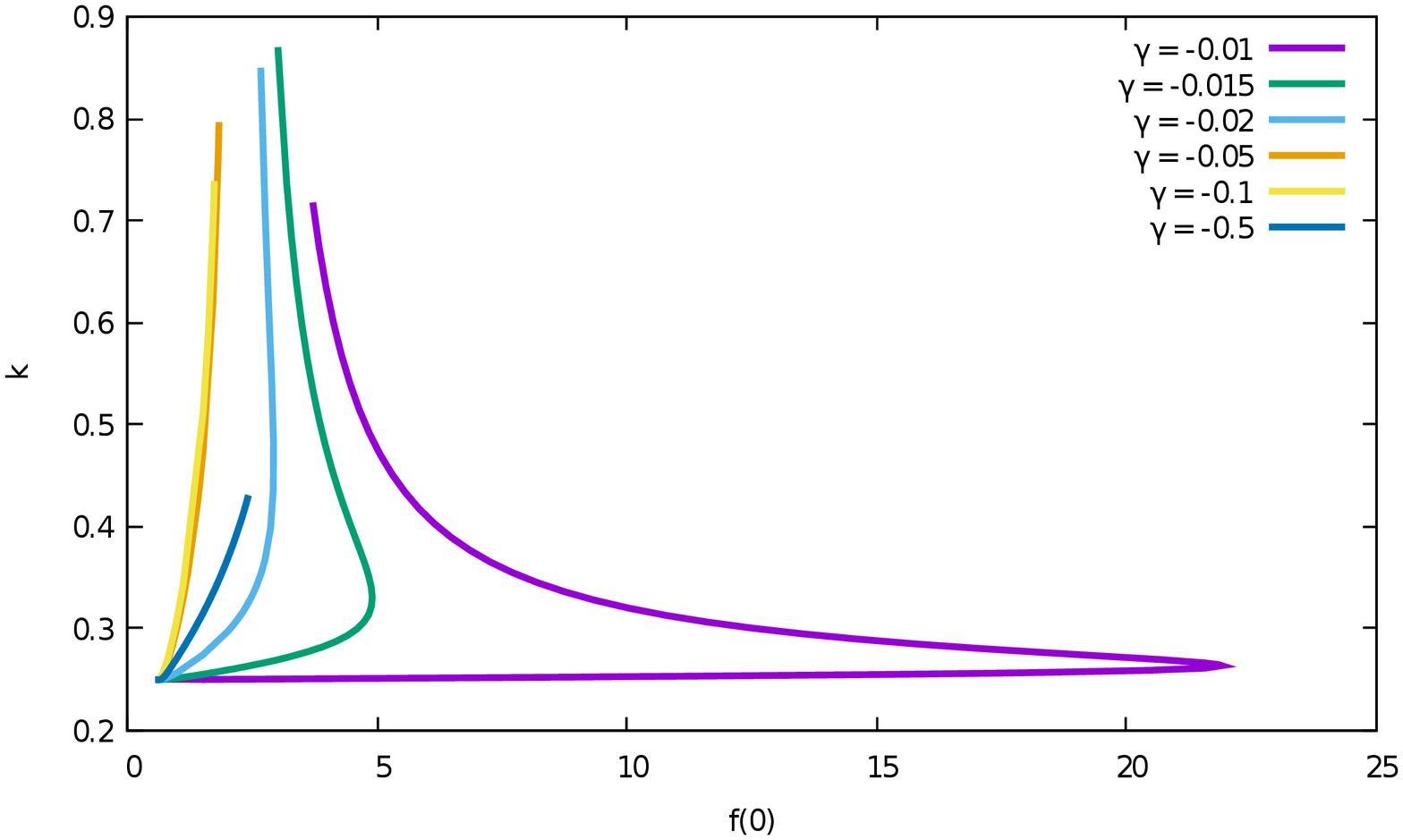}}
\end{center}
\caption{Left: The surface gravity $\kappa$ versus $f(0)$ for several values of $\gamma$ and $Q=0$.
Right: Idem for the solutions with $\gamma < 0$ and $Q=1$. 
\label{kappa}
}
\end{figure}
These solutions do not have a regular limit for $Q \to 0$. For the numerous cases that we considered,
the numerical results indicate that both parameters $\phi(0), a(0)$ tends to zero for 
$Q \to Q_{c}$ and we found the critical value of the charge parameter to be typically  $Q_{c} \sim 0.7$. 
All wormholes of this type that we constructed have $\phi(r)>0$ and $\phi'(r)<0$; as a consequence the 
scalar charge $D$ is negative (contrasting with the  $\gamma > 0$ solutions). 

\begin{figure}
\begin{center}
{\label{non_rot_cc_1}\includegraphics[width=5cm, angle= -90]{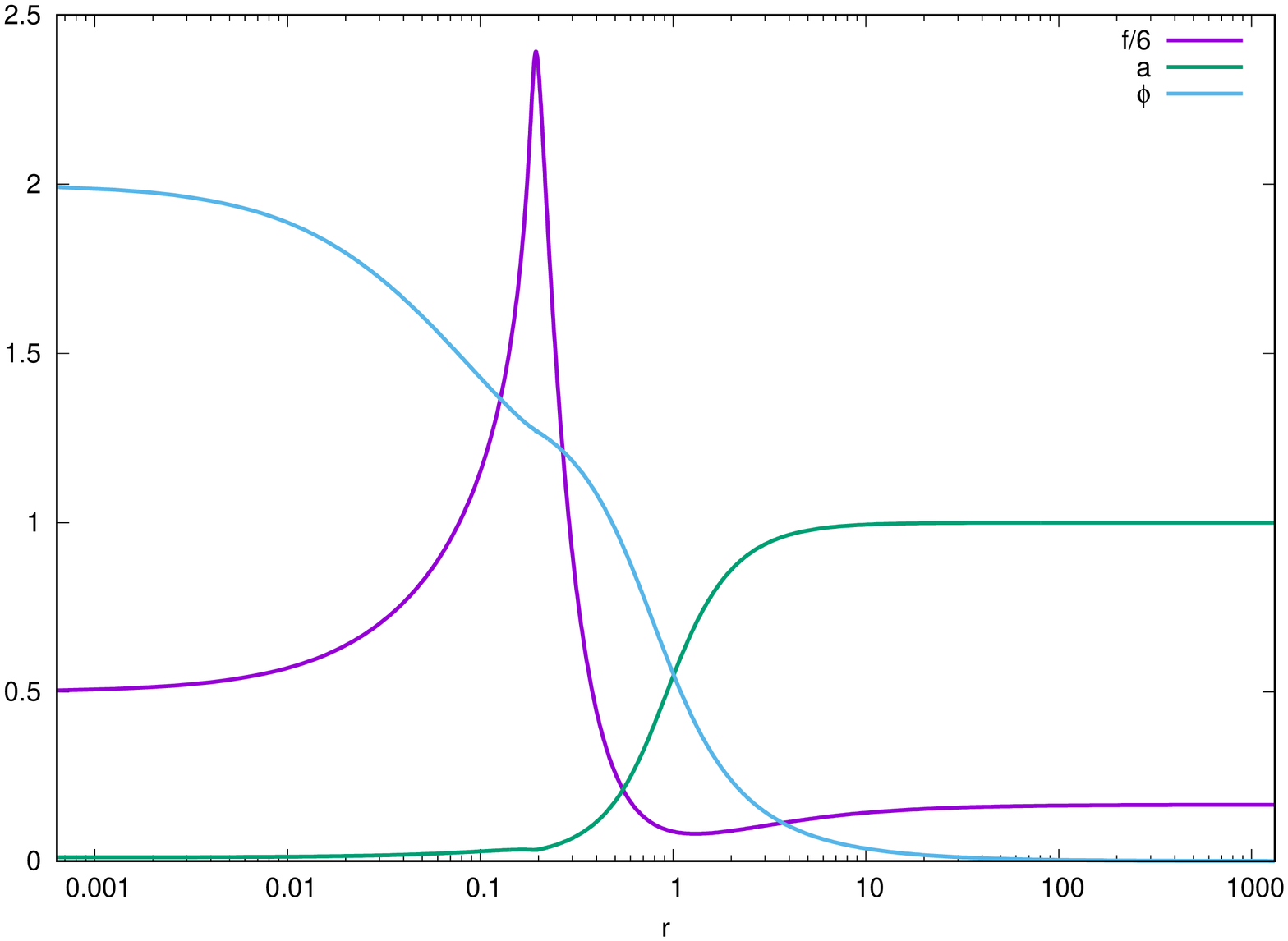}}
{\label{non_rot_cc_2}\includegraphics[width=5cm, angle= -90]{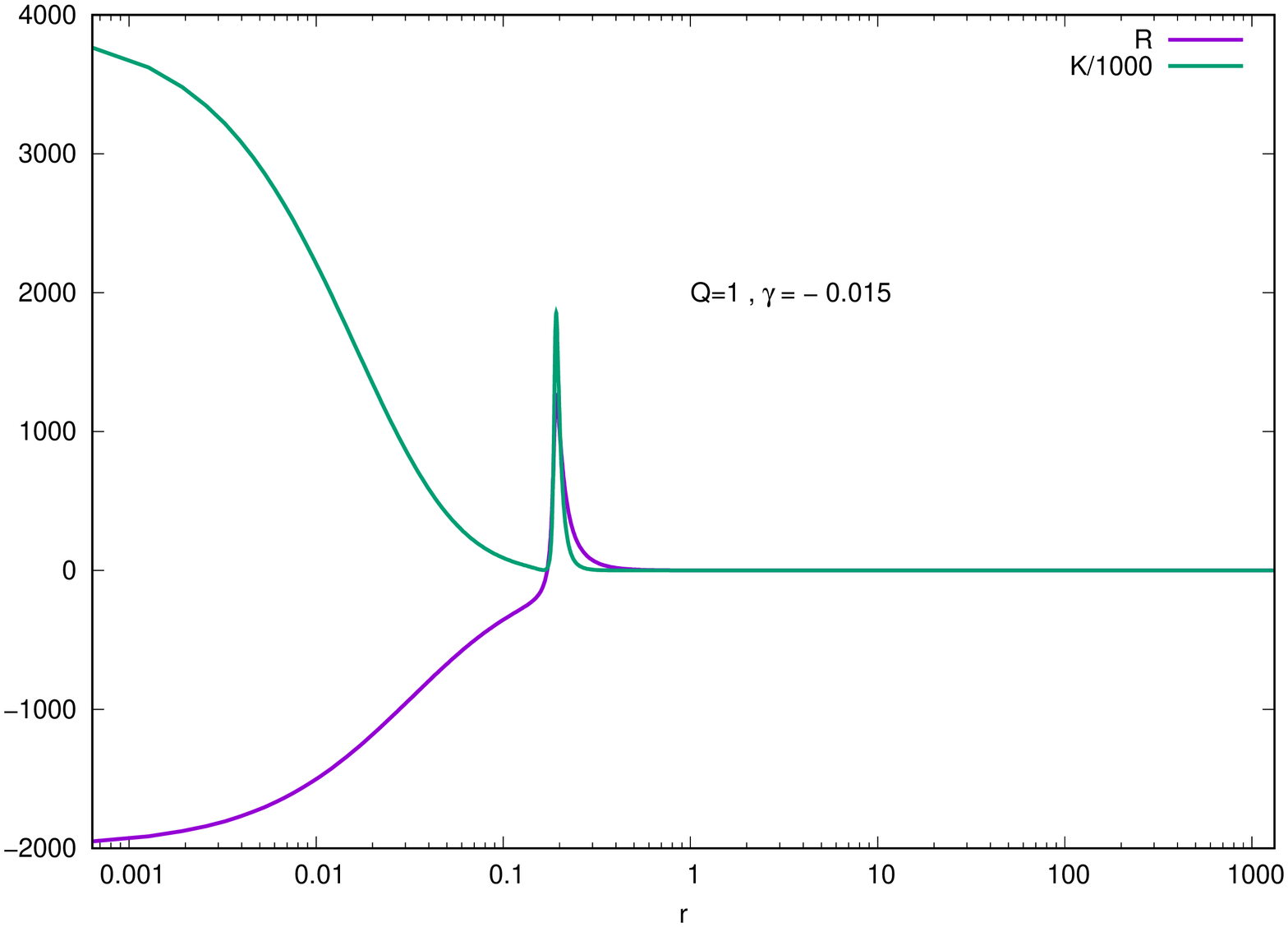}}
\end{center}
\caption{Left: Profile of $f,a,\phi$ for the solution with $f(0)=3$, $\gamma=-0.015$ and $Q=1$.
Right: The corresponding Ricci and Kretschmann invariants.
\label{profile_ga2_m0015}
}
\end{figure} 
The understanding of the full pattern of these solutions is quite demanding. 
We obtained families of wormholes  for several values of $Q$ but, for definiteness, 
we discuss the results for the case $Q=1$; they are  illustrated by Fig.\ref{data_ga2_neg_0}.
\\
For a fixed negative value of $\gamma$  it turns out that wormholes exist for $f(0) \in [f_a , f_b]$
where $f_a$,$f_b$ depend on $\gamma$. 
The minimal  $f_a$ depends only a little from $\gamma$~: we find $f_a \sim 0.55$.
In this limit  the solutions approach a singular configuration as
the parameter $a(0)$ approaches zero while the scalar field approaches uniformly the null function. 
\\
The evolution of the solutions obtained when the parameter $f(0)$ is increases is more involved and depends
strongly of the magnitude of $|\gamma|$. 
As illustrated by Fig.\ref{data_ga2_neg_0} two scenarii clearly occur~:
\begin{itemize}
\item For $|\gamma| \ll 1$,  solutions can be constructed up to a maximal value $f(0) = f_b$, then another branch
of solutions exist. The second branch,  back bending from the main one, 
 terminates at an intermediate value, say $f_c$ with $f_a < f_c < f_b$
in a configuration presenting a singularity at an intermediate value of $r$. The profile of a solution
close to the critical value is presented on Fig. \ref{profile_ga2_m0015}; the Ricci and Kretschmann invariants
reveals  the existence of a singularity at a finite radius. 
Two solutions coexist for $f(0) \in [f_c,f_b]$  and have clearly
different masses (see the lower part of the figure).
\item For $|\gamma| > 0.035 $ only one branch of solutions exist,  
stopping for some $f(0)=f_b$. Again, the limiting configuration presents a singularity
at some intermediate radius.
\end{itemize}
The dependance of $\phi(0)$ on the parameter $f(0)$ is shown on the upper part of
 Fig. \ref{data_ga2_neg_0} for several values of $\gamma$. 
The  lower part of the figure shows the corresponding mass. 
The extend  of the solution in the parameter $f(0)$ decreases progressively while increasing $|\gamma|$.
The solutions reported on the figure are for $|\gamma| \leq 0.5$ but  solutions exist up to $\gamma \sim -5.0$.
\\
Interestingly, the range of the parameter $f(0)$ for the solutions discussed in this section is much
larger that for solutions available for $\gamma > 0$. In particular, charged wormholes can be constructed
for $f(0) \gg 1$, implying that their radius of curvature at the throat, $R_0 = r_0/f(0)$
can be very small; this contrasts
with uncharged  whormholes that have $R_0 \geq 1$. Finally, the surface gravity 
corresponding to the solutions the solutions of Figs. 1 and 2.
is shown on Fig. \ref{kappa} (left and right side respectively).
\\ 
\subsection{Mixed coupling} 
\begin{figure}[h!]
\begin{center}
{\label{non_rot_cc_1}\includegraphics[width=8cm]{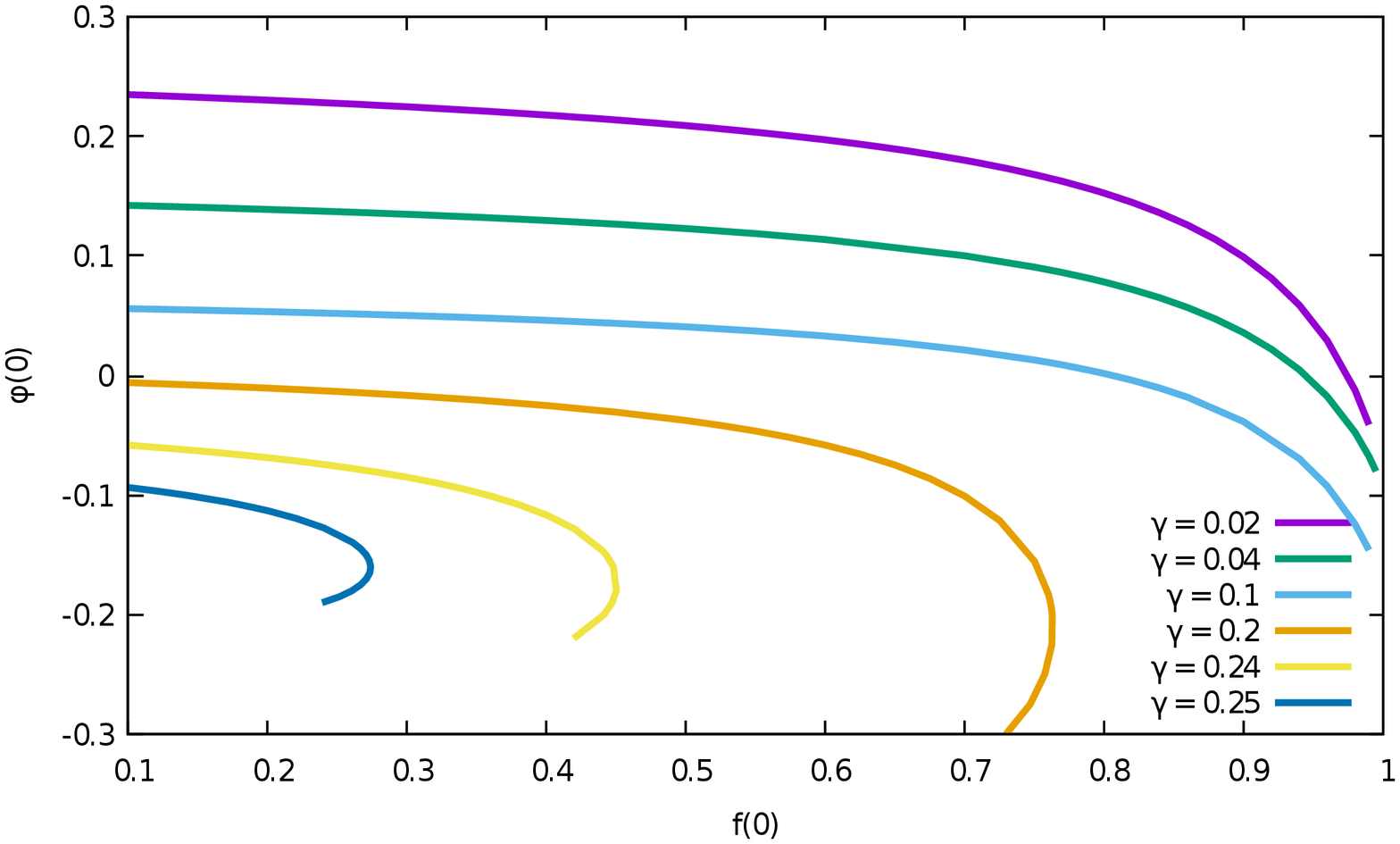}}
{\label{non_rot_cc_2}\includegraphics[width=8cm]{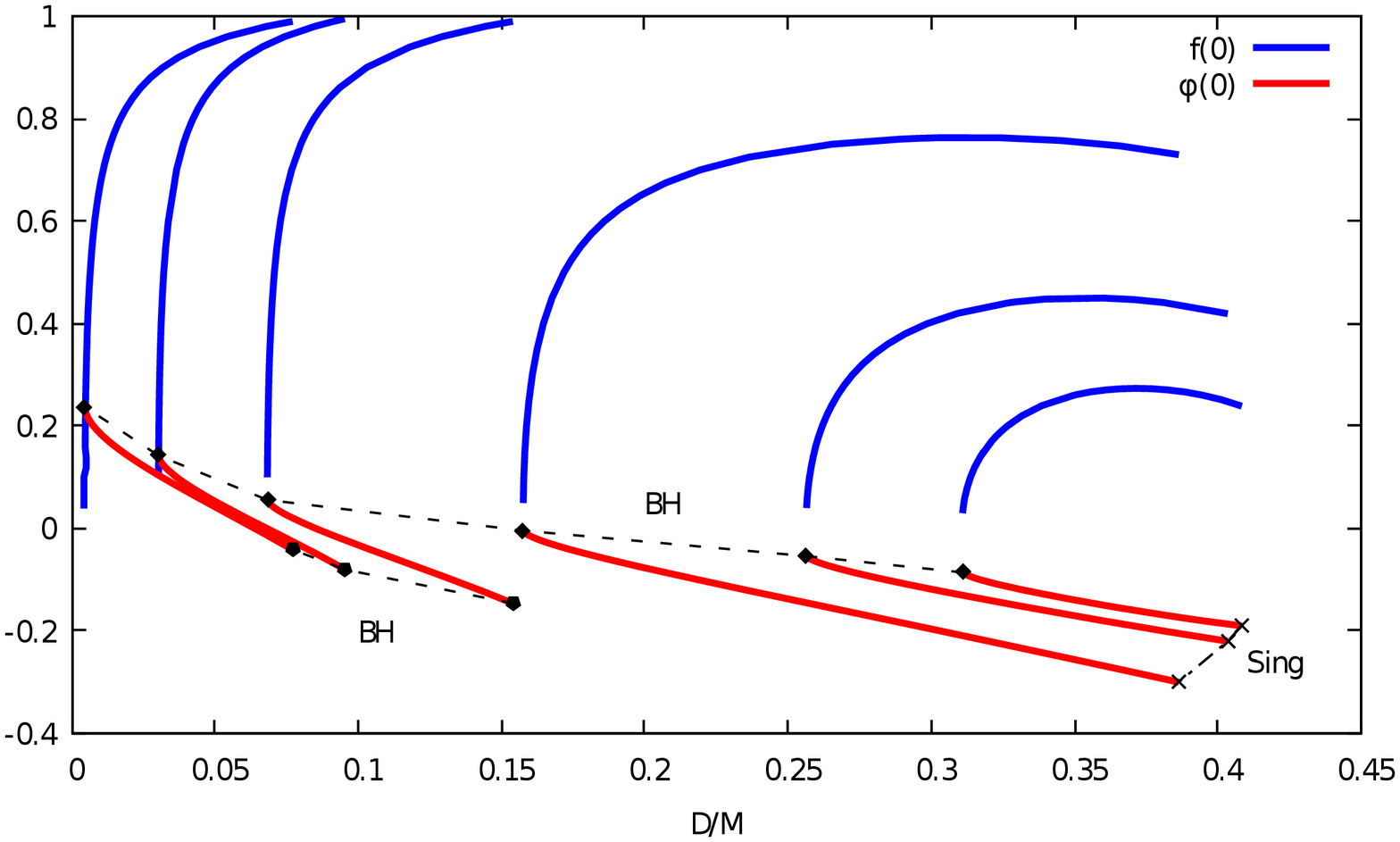}}
\end{center}
\caption{Left: The values $\phi(0)$ as function of $f(0)$
 for several values of $\gamma$ and $\alpha=0.02$. Right: The values $f(0)$ and $\phi(0)$ versus $D/M$;
The curves from the left to the right are for $\gamma = 0.02, 0.04, 0.1, 0.2, 0.25, 0.26$ .
\label{fig_ga_02}
}
\end{figure}

\begin{figure}[h!]
\begin{center}
{\label{non_rot_cc_1}\includegraphics[width=8.0cm]{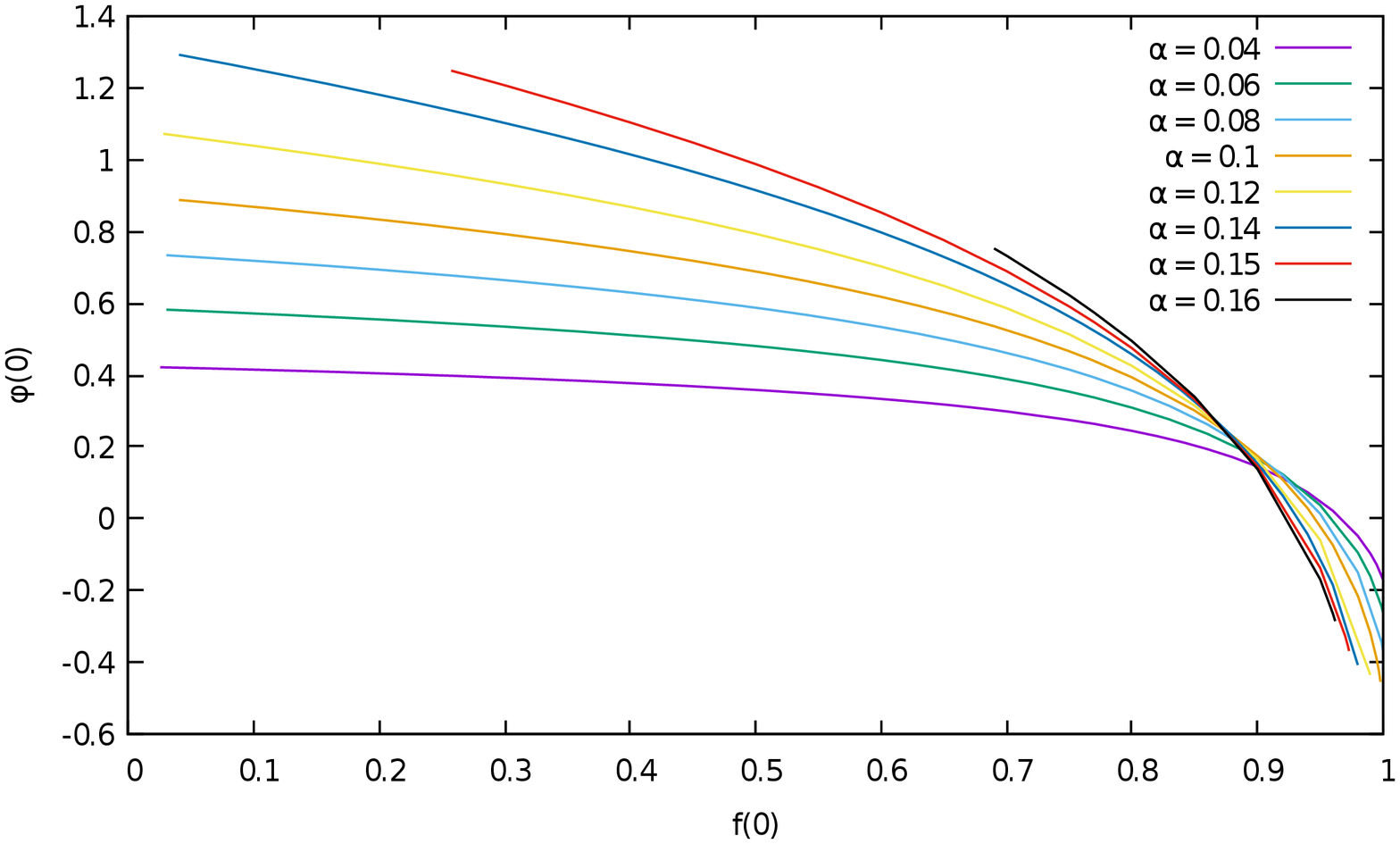}}
{\label{non_rot_cc_2}\includegraphics[width=8.0cm]{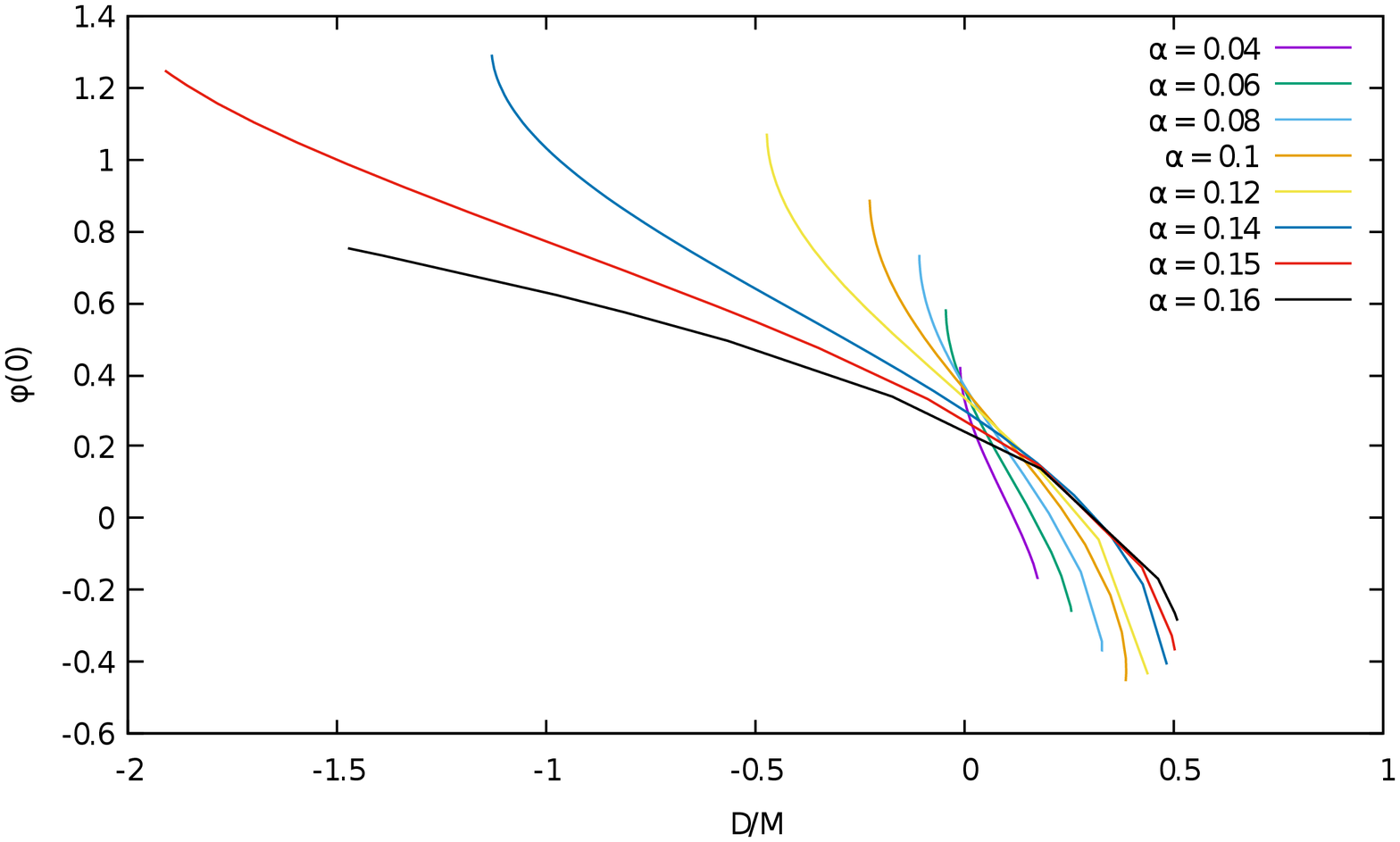}}
\end{center}
\caption{Left: The value $\phi(0)$ as function of $f(0)$ for $\gamma = 0.1$ and several values of $\alpha$.
Right: Idem for the dependance of $\phi(0)$ on the ratio $D/M$.
\label{mixed_bis}
}
\end{figure}

\begin{figure}[h!]
\begin{center}
{\label{non_rot_cc_1}\includegraphics[width=8.0cm]{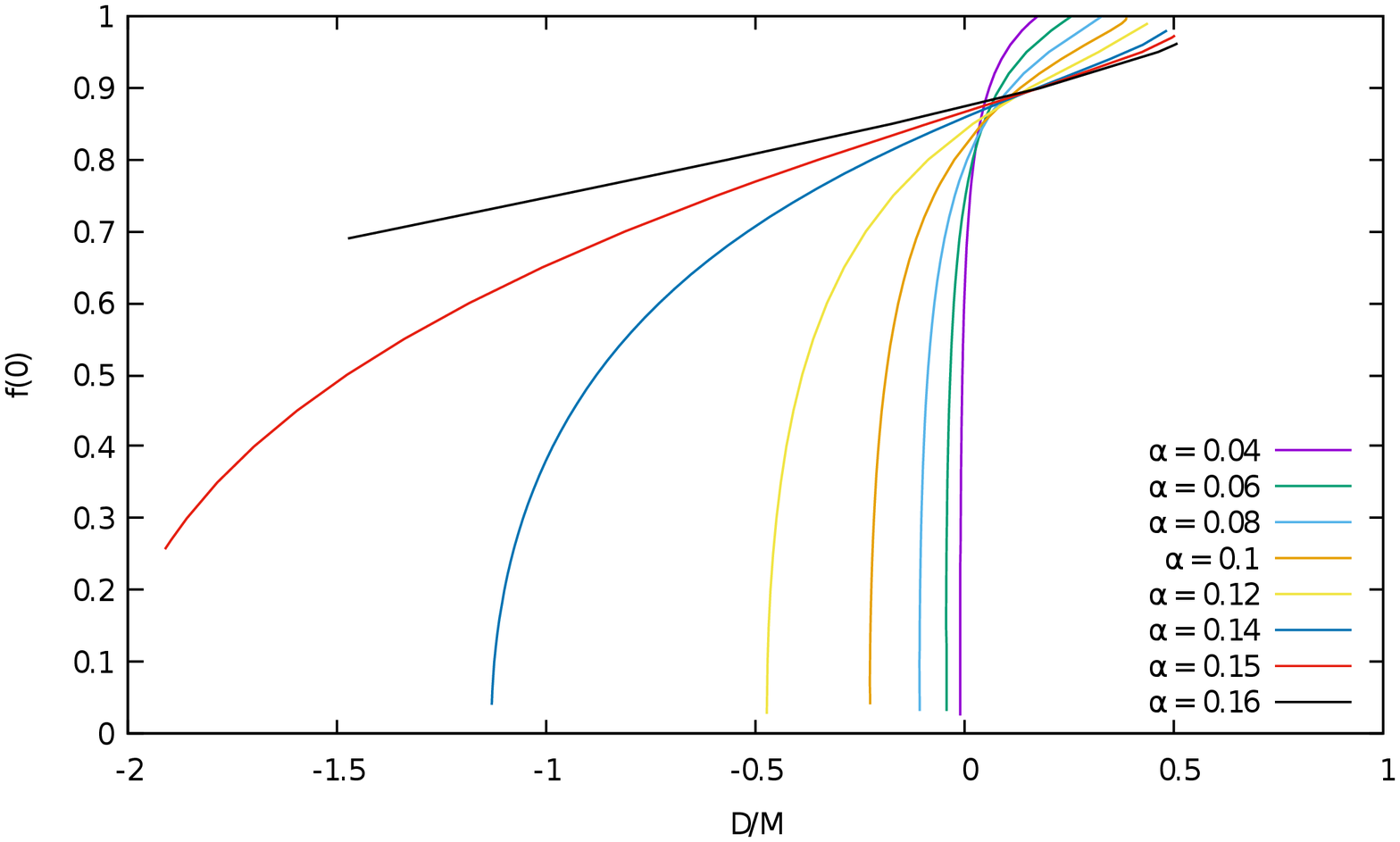}}
{\label{non_rot_cc_2}\includegraphics[width=8.0cm]{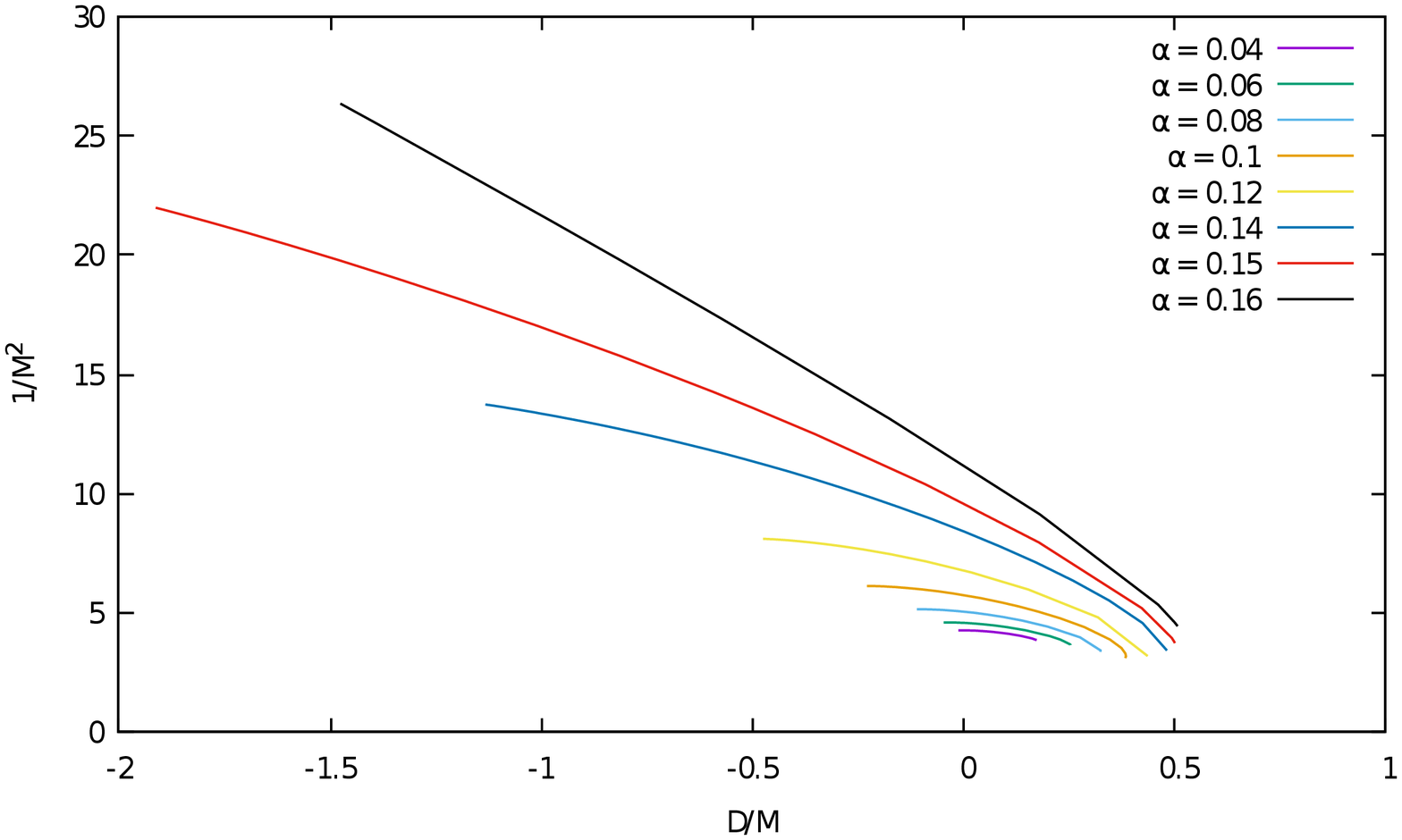}}
\end{center}
\caption{The values $\phi(0)$ (left) and of $M^2$ (right) as function of $D/M$ for $\gamma = 0.1$ and several values of $\alpha$.
\label{mixed_ter}
}
\end{figure}

We now discuss the solutions available with the mixed coupling. 
For simplicity, we limit to the uncharged case. We found no solution for $\gamma \leq 0$; 
therefore the relevant range of parameters is 
 $\alpha \geq 0$, $\gamma \geq 0$. We will  sketch the influence of these parameters on the pattern of solutions
 by presenting results respectively for 
 $\alpha$ fixed and  $\gamma$ varying   and for  $\gamma$ fixed and $\alpha$ varying. 
\\
{\bf Case $\alpha$ fixed, $\gamma$ varying.}
\\
Perhaps one of the striking features in the case $\alpha > 0$ is the fact that wormholes exist  
for $\gamma > 0$~: contrasting  with  the pure quadratic case  
(see  section 3.1) there is no threshold in the coupling constant $\gamma$. 
We therefore put the emphasis on families of solutions occuring for $\alpha \ll 1$. 
Some new features of the solutions are sketched on Fig. \ref{fig_ga_02}
for several values  of $\gamma$ and where we set for definiteness $\alpha = 0.02$.
A single branch of solutions occurs for $0 < \gamma < 0.175]$; 
it is labeled by $f(0)$ (see left side of Fig. \ref{fig_ga_02}) and extend for $f(0) \in ]0,1[$.
The central value of the scalar field $\phi(0)$ is  positive or negative along  the branch.
For $\gamma > 0.175$ two families of wormholes exist and coincide at a maximal value of $f(0)$ with $f(0) < 1$. 
All  these solutions have a positive scalar charge: $D > 0$ as seen on the right side of Fig. \ref{fig_ga_02}.
\\
{\bf Case $\gamma$ fixed, $\alpha$ varying.}
\\
We finally analyze  the influence of the increase of the linear coupling constant $\alpha$
on a solution with  a fixed $\gamma$;  for definiteness we concentrate on 
solutions corresponding to $\gamma = 0.1$ (remember: they have no $\alpha=0$-limit).
The deformation of the physical data by the increase of $\alpha$ is illustrated by Figs. \ref{mixed_bis}, \ref{mixed_ter}.
We see in particular that the increase of $\alpha$ allows for solutions with positive
central density of the scalar field~:
solutions with $\phi(0) > 1$ typically exist while all solutions for $\alpha = 0$ have $\phi(0) < 0$. 
Plotting the data as function of the dimensionless parameter $D/M$ also reveals new features.
It turns out that 
the interval of $D/M$ where solutions are available considerably while $\alpha$ increases.
A large fraction of wormholes with mixed coupling have a negative scalar charge $D$. 
Typical  solutions have  $D/M \in [-1, 0.5]$; again contrasting with all $\alpha=0$ solutions.
\section{Summary}
The Eintein-Hilbert-Maxwell-Klein-Gordon action considered in this paper is extended by a non-minimal
interaction involving the Gauss-Bonnet term coupled to a specific function $H(\phi)$ of the scalar field $\phi$.
The choice $H(\phi) = \alpha \phi + \gamma \phi^{2}$ is motivated  by the fact that
 hairy black holes are known to exist in the two limits $\alpha=0$ and $\gamma=0$. 
It is therefore natural to emphasize the existence of wormholes separately in both cases and to further
study how these solutions evolve  in the mixed case. 
It was first demonstrated numerically that wormholes appear spontaneously in the case $\alpha=0$ at a critical value of $\gamma$.
By continuity it was then shown  that,  for a large domain of the 
coupling constants $\alpha$ and $\gamma$,   
the model possesses wormholes solutions crucially supported by the non-trivial scalar field. 
Let us stress that this scalar field has a conventional kinetic term in the Lagrangian.

The influence of the electromagnetic field, characterized by the electric charge $Q$, was also taken into account
leading to families of charged wormholes solutions.
Thereby, new classes of  solutions have been constructed which present different features from the wormholes constructed
with the help of the Gauss-Bonnet term. Namely: (i) they present very small curvature radius at the throat, (ii)
they have no smooth limit for $Q \to 0$, (iii) they exist with both signs of the scalar field at the throat.
Nevertheless, work is still needed to examine further properties of these new wormholes, namely their stability and their analytic continuation
in the $r < 0$ region.

\pagebreak


 \end{document}